\documentclass[11pt,a4paper]{article}
\pdfoutput=1
\usepackage{jcappub}
\usepackage{graphicx}
\usepackage{subfig}

\def\apj{{\it Astrophys.~J.}}
\def\apjs{{\it Astrophys.~J. Suppl.}}

\def\apjl{{\it Astrophys.~J.~Lett.}}
\def\prd{{\it Phys.~Rev.~D}}
\def\prl{{\it Phys.~Rev.~Lett.}}
\def\plb{{\it Phys.~Letts.~B}}
\def\mnras{{\it Mon.~Not. Roy.~Astr.~Soc.}}
\def\ijmpd{{\it Int.~J.~ Mod. Phys. D}}

\def\AnA{{\it Astron. Astrophys.}}

\def\ARAnA{{\it Ann. Rev. Astron. Astrophys.}}

\title{Is dark energy evolving?}
\author[a]{Remya Nair,}
\author[a]{Sanjay Jhingan}

\affiliation[a]{Centre for Theoretical Physics,\\ Jamia Millia
Islamia, New Delhi 110025, India}

\emailAdd{remya$_{-}$phy@yahoo.com}
\emailAdd{sanjay.jhingan@gmail.com}

\abstract{We look for evidence for the evolution in dark energy
density by employing Principal Component Analysis (PCA). Distance
redshift data from supernovae and baryon acoustic oscillations (BAO)
along with WMAP7 distance priors are used to put constraints on
curvature parameter $\Omega_k$ and dark energy parameters. The data
sets are consistent with a flat Universe. The constraints on the
dark energy evolution parameters obtained from supernovae (including
CMB distance priors) are consistent with a flat $\Lambda$CDM
Universe. On the other hand, in the parameter estimates obtained from
the addition of BAO data the second principal component,
which characterize a non-constant contribution from dark energy, is
non-zero at 1$\sigma$. This could be a systematic effect and future
BAO data holds key to making more robust claims.}

\keywords{cosmic acceleration, supernovae, baryon
acoustic oscillations, dark energy}

\begin{document}

\maketitle

\section{Introduction}
The late time acceleration of the Universe as suggested by type Ia
Supernovae (SNeIa) \cite{exp} observations is now confirmed by
various different probes \cite{DHwein}. Exploring the physics behind
the cosmic acceleration is the focus of next generation surveys.
The discovery of an accelerating phase of the Universe poses a very
pressing question: what is the driving force responsible for this
acceleration and what are its properties. There are two possible
ways explored by theorists to answer this question: modify gravity
on large scales or invoke a non-standard quantity in the framework
of general relativity, commonly termed `dark energy' which has
negative pressure \cite{sah}. Many theoretical models have been
proposed for dark energy, the cosmological constant $\Lambda$ being
the simplest, and evolving dark energy scenario explained by using
scalar field models \cite{samik}. An alternative approach which is
complimentary to dark energy model building, is deriving the dark
energy properties from the data. The dark energy models are often
characterized by the equation of state parameter $w=p/\rho$.
For the case of the cosmological constant $w=-1$, and for dynamical
models $w$ is a variable. One of the main targets of future
Cosmological surveys is to find constraints on the equation of
state and also on its evolution (if any).

To quantify the possible dependence of the
dark energy properties on redshift, one can either parameterize them
or reconstruct them from the data in a non-parametric way. The equation of state
is often parameterized as $w = w_0 + w_a(1-a)$. There have been attempts to study
the accelerated expansion of the Universe using other kinematic variables
like the Hubble parameter $H(z)$, the deceleration parameter $q(z)$,
or the jerk parameter $j(z)$ which are all constructed from
derivatives of the scale factor $a$ \cite{kin}. Direct parameterization has an 
advantage that one can physically interpret the result easily and quantify the
dark energy and other kinematical properties with a few numbers \cite{lindr, kin}.
But this approach can introduce bias in the analysis since the result will
always depend on the form of the parameterization chosen
\cite{bruce1}. For this reason non-parametric methods have received
a lot of attention in the past few years.

Huterer and Starkman were the first to propose the use of Principal
component analysis to obtain information about the properties of
dark energy \cite{hut}. They assumed a fiducial survey with 3000
SNeIa distributed uniformly over a redshift range $0 \leq z \leq
1.7$. Using this simulated data set they derived the best determined
weight functions for the equation of state $w(z)$. Later Shapiro and
Turner used supernovae measurements to analyse the acceleration
history assuming a flat spacetime that is homogeneous and isotropic
on large scales. Using principal component analysis they found very
strong (5$\sigma$) evidence for a period of acceleration and strong
evidence that the acceleration has not been constant \cite{shap}.
Wang and Tegmark proposed a method for measuring the expansion
history of the Universe in uncorrelated redshift bins \cite{wang}.
Zunckel and Trotta used a maximum entropy method based on Bayesian
framework to reconstruct the equation of state of dark energy
\cite{zunc}. Sarkar et al., constrained the equation of state of
dark energy by using uncorrelated binned estimate and showed that
more than three independent parameters of the equation of state can
be obtained from future dark energy surveys to an accuracy better
than 10$\%$ \cite{sark}. Gaussian process (GP) modelling has also
been used to reconstruct the dark energy equation of state \cite{marina}. Holsclaw
et al., used GP modelling and showed that the non-trivial behaviour
of $w$ as a function of $z$ can be extracted from future data and they apply their method on SNeIa data to
reconstruct the history of the dark energy equation of state out to
redshift $z=1.5$ \cite{hols1}. Ishida et al., used PCA to reconstruct the
expansion rate of the universe with SNeIa data \cite{ishida}.

The aim of this work is to find evidence for evolution in the dark
energy density. We have used PCA to find constraints on the amplitudes 
of modes of the dark
energy density as a function of redshift \cite{jason}. The plan of
the paper is as follows: in section 2 we briefly review the
parameter degeneracy between the curvature and dark energy and we
introduce the data sets used in this work and the methodology in
section 3. In section 4 we present our main results and discuss our
results in section 5.

\section{Parameter constraints and degeneracy}

Finding constraints on the dark energy evolution parameters is
complicated by the `geometric degeneracy' between dark energy and
curvature (see \cite{degn} and references therein). It is a common
practice to assume spatial flatness to find constraints on dark
energy parameters, and a simplified dark energy model is assumed
when constraining curvature. For example, if curvature is a free
parameter, the equation of state of dark energy is either assumed to
be a constant or is parameterized with some simple form:
$w(a)=w_0+w_a(1-a)$. The degeneracy arises, as the observed
distances depend both on the expansion history of the Universe and
the curvature. As a result it is not possible to constrain both the
curvature and dark energy parameters. As also discussed by Mortonson
\cite{Mort} (eq.(17) in the paper), if one uses only distance
measurements, then for any value of the spatial curvature, one can
derive some dark energy evolution to satisfy the observations.
Mortonson used growth data to remove this degeneracy, since unlike
the distance data, growth data depends only on the expansion rate.
In an attempt to demonstrate that including the curvature as a free
parameter is imperative to understand the dark energy evolution,
Clarkson et al., showed that the assumption of a flat universe leads
to large errors in the reconstruction of the dark energy equation of
state even if the true cosmic curvature is very small \cite{clark}.
Similarly Shafieloo and Linder analysed the degeneracies that arise
in the distance-redshift relation when there is no a priori
restriction on the equation of state of dark energy \cite{shaf} and
they found that large variation in the parameters are allowed when
using only distance measurements.

In this work we assume a homogeneous and isotropic Universe
described by the FLRW metric. We keep $\Omega_k$ as a free parameter
along with the dark energy parameters to be constrained by the data.
\section{Methodology and data sets used}
\subsection{Data sets}
The dark energy evolution effects the expansion rate of the Universe
and hence the distances on cosmic scales. In this work we have
used distance-redshift data from SNeIa and BAO measurements. Listed below are the publicly available data
sets used here:

\begin{itemize}
\item We use SNeIa Union2.1 sample as described in \cite{suz} to estimate the luminosity
distance. This sample contains 580 supernovae spanning the redshift range $0.015<z<1.414$.
\item BAO data from different galaxy cluster
surveys - SDSS ($z$=0.2, 0.35), 6dFGS ($z$=0.106), WiggleZ ($z$=0.44,
0.6, 0.73) and BOSS ($z$=0.57) \cite{bl1,perc,beut,boss1}.
\end{itemize}
In addition to the above data sets we also use the WMAP7 distance
priors \cite{wmap}. The physics at the decoupling epoch ($z_*$)
affects the amplitude of the acoustic peaks. The evolution of
the Universe between now and $z_*$ effects the angular diameter
distance out to decoupling epoch, and hence the locations of the peaks. This
information is encoded in the `acoustic scale' $l_A,$ and the `shift
parameter' $R$ derived from the power spectrum of cosmic microwave
background.
\subsection{Methodology}
The dark energy contribution to the expansion rate is expressed as a
sum of two components. The contribution from the first term is
constant across the redshift range. The second component
corresponds to a variation in dark energy density and hence its
contribution varies across the redshift range. We express the second
term as a binned expansion, i.e. we divide the redshift range of the
data in bins, and given a complete basis set $\{e_i\}$, the contribution from
the second term can be written in terms of the basis vectors.
If the redshift range is divided in, say, $N$ number of bins then
every element in the $N\times$1 basis vector can be associated with
a redshift bin. The continuum limit is reached as $N\rightarrow
\infty$. Thus the dark energy density is expanded as:
\begin{equation}
\rho _{\Lambda}(z)=\rho _c \left(\alpha _0+ \sum_{i=1}^N \alpha _i e_i (z)\right).
\end{equation}
Here $\alpha _0$ specifies the contribution which is constant in
redshift and the coefficients $\alpha_i$ (for $1 \leq i \leq N$)
specify the evolution in dark energy density. $\alpha$'s thus
determine the dark energy density upto an overall constant $\rho _c$
which is the critical energy density today. For $\alpha _0 = \Omega
_\Lambda$ and all other $\alpha _i$'s = 0, we recover the standard
$\Lambda$CDM case. The choice of the basis is arbitrary. We chose
the basis vectors so that $e_i(z)$=1 in the $i^{th}$ redshift bin
and zero otherwise (i.e. we chose the $N$ $\times$ $N$ identity matrix to be the initial basis). In this work the results are produced with $N$=50 bins.

{\bf SNeIa}: For the standard FLRW metric the luminosity distance is given by
\begin{equation}
d_L = \frac{c(1+z)}{\sqrt{|\Omega _k| H_0^2}} ~
S_k\left(\sqrt{|\Omega _k| H_0^2} \int _0 ^z
\frac{dz'}{H(z')}\right),
\label{dl}
\end{equation}
where $S_k(x)$ is equal to $\sin{x}$, $x$, or $\sinh{x}$
corresponding to closed, flat and open Universe and the expansion
rate of the Universe is:
\begin{equation}
H(z)=H_0 [\Omega_m (1+z)^3+\Omega_k (1+z)^2+\Omega_r
(1+z)^4+\Omega_\Lambda f(z)]^{1/2}. \label{hz}
\end{equation}
Here $H_0$ is the value of the Hubble parameter at present and
$f(z)$ captures the form of dark energy evolution. $\Omega_i$ is the
density parameter, defined as $\Omega_i=\rho_i(z=0)/\rho_c$ and
$\Omega_m$, $\Omega_k$, $\Omega_r$ and $\Omega_\Lambda$ refer to the
contribution in the energy density at the present epoch from matter,
curvature, radiation and dark energy respectively and they add to
give unity i.e.
\begin{equation}
\Omega_m + \Omega_k + \Omega_r + \Omega_\Lambda =1.
\end{equation}
The value of $\Omega_r$ can be neglected in the redshift range
corresponding to the supernovae and BAO data. The distance
modulus $\mu = m-M $, which is obtained from the Union2.1
compilation can be derived from the luminosity distance as
\begin{equation}
\mu _{th} = 5 \log _{10} \frac{d_{L}}{Mpc}+25.
\end{equation}
Here $m$ and $M$ are the apparent and absolute magnitudes
respectively. The cosmological parameters are estimated by
minimizing the chi-squared merit function:
\begin{equation}
\chi_{_{Union2}}^2  =\sum_{i=1}^{580}
\frac{(\mu ^{th}(z_{i},p)-\mu ^{obs}(z_{i}))^2}{\sigma^{2}_{\mu _{i}}},
\end{equation}
where $p$ is the set of parameters ($\alpha _i , \Omega _k, h$).

{\bf BAO}: Galaxy cluster surveys
provide measurements of an angle-averaged distance $D_{V}$
\begin{equation}
D_{V}=\left(\frac{cz(1+z)^{2}d_{A}^{2}}{H(z)}\right)^{\frac{1}{3}},
\label{dv}
\end{equation}
or the distilled parameter $d_z = r_{s}(z_d)/D_{V}$. Here $d_{A}$ is the angular diameter distance which is theoretically given by $d_L/(1+z)^2$ and $r_s(z_{d})$ given by
\begin{equation}
r_s(z)=\frac{c}{\sqrt{3}} \int _0 ^{1/1+z} \frac{da}{a^2 H(a) \sqrt{1+(3\Omega _b/4\Omega_ {\gamma})a}}
\end{equation}
is the characteristic scale determined by the comoving sound horizon at an
epoch $z_d$ slightly after decoupling. This epoch is measured by
CMB anisotropy data. We determine $r_s(z_d)$ using the fitting formula given by Percival et al. \cite{perc}
\begin{equation}
r_s(z_d)=153.5 \left(\frac{\Omega _b h^2}{0.02273} \right)^{-0.134} \left(\frac{\Omega _m h^2}{0.1326} \right)^{-0.255} Mpc.
\end{equation}
Here $\Omega_b$ is the baryon density and $h$ is defined as $h=H_0/100$ km/s/Mpc. We have 7 BAO points
in the redshift range $0.106 \leq z \leq 0.73$. Since some of these points are correlated we use the
corresponding covariance matrix $C$ and the chi-squared for this data is:
\begin{equation}
\chi_{_{BAO}}^2  = (D)^T C^{-1}D
\end{equation}
where $D=d_z^{obs}(z)-d_z^{th}(z,p)$ is a column matrix for the seven data points and $p$
is the set of parameters ($\alpha _i , \Omega _k, h$).

{\bf CMB distance prior}: The `acoustic scale' $l_A$, the CMB `shift parameter'
$R$, and the redshift to decoupling $z_*$ mentioned earlier are defined as \cite{wmap}:
\begin{eqnarray}
l_A &\equiv & (1 + z_*) \frac{\pi d_A(z_*)}{r_s (z_*)},\\
R(z_*) &\equiv & \frac{\Omega _m H_0^2}{c} (1 + z_*) d_A(z_*),\\
z_* &=& 1048[1 + 0.00124(\Omega _b h^2)−0.738][1 + g_1(\Omega _m h^2)g_2].
\end{eqnarray}
where $g_1$ and $g_2$ are given by
\begin{eqnarray}
g_1 &=& \frac{0.0783(\Omega _b h^2)^{−0.238}}{1 + 39.5(\Omega _b h^2)^{0.763}} \;,\\
g_2 &=& \frac{0.560}{1 + 21.1(\Omega _b h^2)^{1.81}} \;,
\end{eqnarray}
and the chi-squared can be written as:
\begin{equation}
\chi_{_{CMB}}^2  = (D)^T C^{-1}D\;.
\end{equation}
Here $D$=$(l_A^{th}(p) - l_A^{obs}$, $R^{th}(p)-R^{obs}$, $z_*^{th}(p)-z_*^{obs})^T$,
$C$ is the corresponding covariance matrix and $p$ is the set of parameters
($\alpha _i , \Omega _k, h, \Omega _b$).

{\bf Fisher Matrix and eigenmodes}: We assume that our data set is
composed of independent observations and is well approximated by a
Gaussian probability density. Now since the measurements are
independent one can take the product of their corresponding
likelihood functions to obtain the combined likelihood function. One
can maximise this likelihood function to find the best fit
parameters or one can minimize the corresponding chi-squared merit
function. The combined chi-squared for all the data sets will be a
sum of chi-squares from individual  measurements
\begin{equation}
\chi_{_{Total}}^2  = \chi_{_{Union2}}^2+\chi_{_{BAO}}^2+\chi_{_{CMB}}^2.
\end{equation}
The cosmological parameters in our analysis are as follows:
curvature parameter $\Omega_k$, the Hubble parameter $h$, dark
energy density parameters $\alpha _i$'s and the baryon density
$\Omega _b$. Thus if we have $N$ bins we have $N+$4 unknown
parameters. From this chi-square one can construct the Fisher
matrix. The Fisher Matrix is defined as the expectation value of the
derivatives of the log of the likelihood $L(\propto
e^{-(\chi^2/2)})$, with respect to the parameters:
\begin{equation}
F_{ij}=-\left \langle \frac{\partial^2 \ln L}{\partial p_i \partial p_j} \right\rangle \;.
\end{equation}
Depending upon the data set under consideration we construct the corresponding
Fisher matrix. For this, one has to chose a particular point $p_0$ in the parameter
space at which the Fisher matrix is evaluated. We chose this point to correspond to the standard $\Lambda$CDM case,
$\alpha_0 = \Omega_{\Lambda}$ and $\alpha_i=0$ for $1 \leq i \leq N$. Once
we have the Fisher matrix, we marginalize over all the parameters other than
$\alpha_i$ ($1 \leq i \leq N$), i.e. $\alpha_0$, $\Omega_k$,
$h$ and $\Omega_b$. We are left with a Fisher matrix of the
parameters that correspond to the evolution in dark energy density. We diagonalize
it so that it can be written as
\begin{equation}
F=W\Lambda W^T,
\end{equation}
where the columns of the orthogonal decorrelation matrix $W$, are
the eigenvectors of $F$ and $\Lambda$ is a diagonal matrix with the
eigenvalues on its diagonal. We can now find a new set of basis
functions which is a linear combination of the old basis function
$e_i$. This new basis set has decorrelated vectors and since these
basis functions are orthonormal and complete we can write the
contribution of dark energy in terms of these uncorrelated basis
vectors $c_i$:
\begin{equation}
\sum_{i=1}^{N} \alpha_i e_i(z)=\sum_{i=1}^{N} \beta_i c_i(z) \;.
\end{equation}
The eigenmatrix $W$ is the Jacobian matrix of the transformation of
one basis to another, so one can construct these new basis functions
$c_i(z)$ as (see \cite{amara} for details)
\begin{equation}
c_i(z)= W e_i(z)
\end{equation}

\begin{figure}
\centering  
{\includegraphics[width=4in]{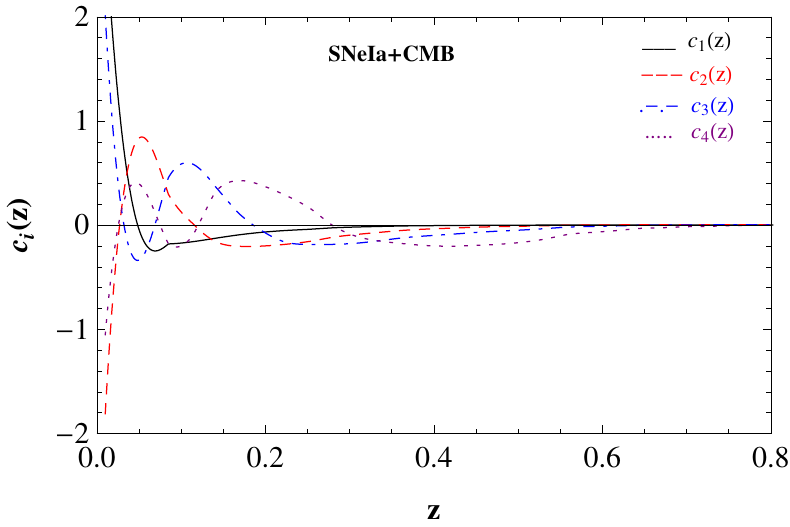}}
\caption{The panel shows the eigenmodes obtained for the
Supernovae+CMB data set. Solid (black), dashed (red), dot-dashed
(blue) and dotted (purple) curves correspond to the first, second,
third and fourth eigenmode respectively.}
\label{sc}
\end{figure}

\begin{figure}
\centering  
\subfloat[Part
1][Dark energy parameters]{\includegraphics[width=3.0in]{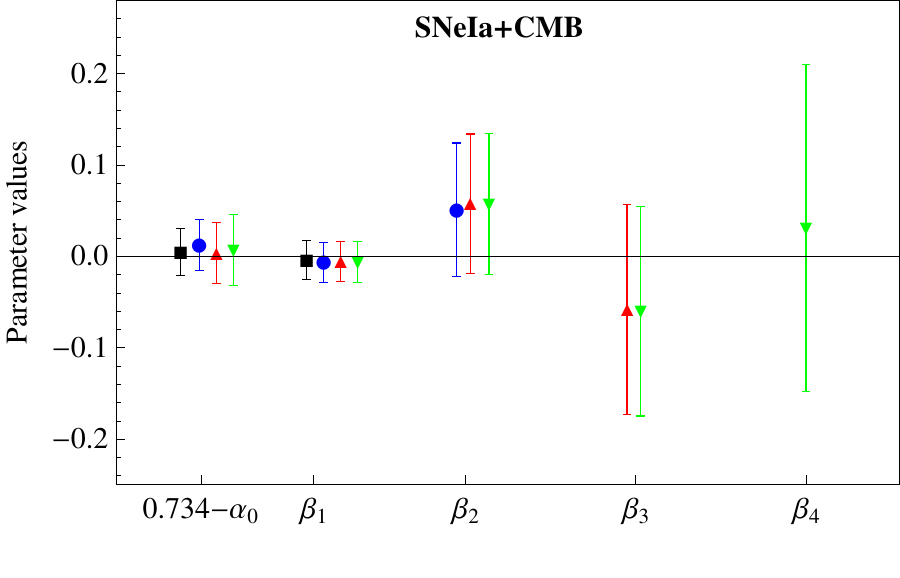}\label{sc1}}
~~~~~\subfloat[Part 1][Curvature parameter]{\includegraphics[width=3.0in]{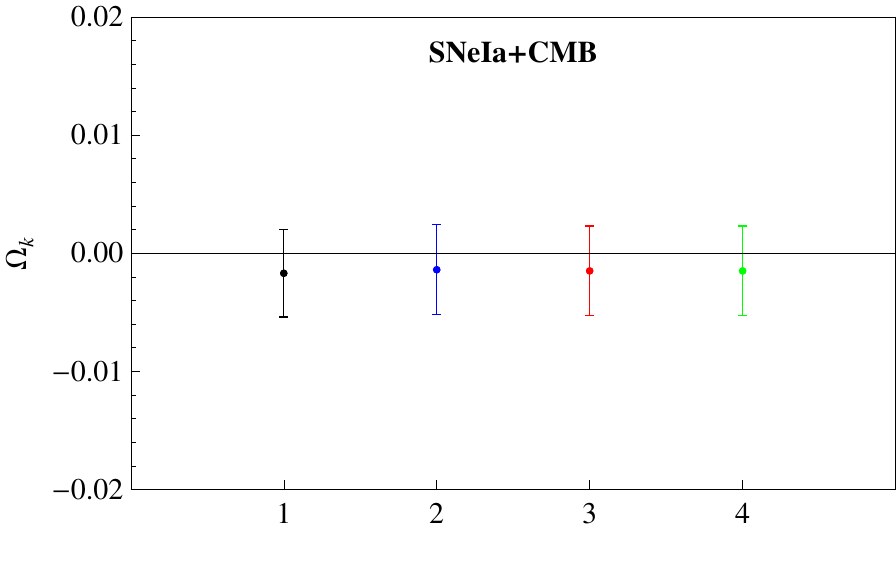} \label{kc2}}
\caption{The left panel shows the
estimates for dark energy parameters. Color scheme
(from left to right): Black/square (Case 1): $\alpha_0$ $\&$ $\beta_1$
allowed to vary, Blue/circle (Case 2): $\alpha_0$, $\beta_1$ $\&$
$\beta_2$ allowed to vary, Red/up arrow (Case 3): $\alpha_0$, $\beta_1$,
$\beta_2$ $\&$ $\beta_3$ allowed to vary, Green/down arrow (Case 4):
$\alpha_0$, $\beta_1$, $\beta_2$, $\beta_3$ $\&$ $\beta_4$
allowed to vary. The right panel shows the curvature parameter estimates.}
\end{figure}

Note that in our case the initial basis set is just the identity
matrix. The advantage of this new basis is that the $\beta_i$ are
uncorrelated, which implies that  any pair of coefficients has a
non-degenerate error ellipse. These eigenmodes are arranged from the
best determined mode to the least determined mode i.e., from the
largest to the smallest eigenvalues (the error on these modes goes
as $\sigma \propto \lambda^{-1/2}$) and then we choose the first few
best determined eigenmodes to reconstruct the dark energy density
and carry out the chi-squared minimization. Also note that the
constant mode $\alpha_0$ is not decorrelated from the rest of the
modes and is unaffected by the diagonalisation of Fisher matrix. We
show the four best determined eigenmodes for the different data set
combinations in the figures below. A non-zero mode amplitude of
these principal components would indicate time evolution in dark
energy.

The main advantage of the PCA is dimensionality reduction. Initially
our parameter space had $N+4$ parameters. Now, depending on the
number of principal components chosen for reconstruction, the
parameter space would be reduced. The number of principal components
to be used for reconstruction depends on how much information are we
willing to discard. If we used all the principal components, we
would have 100$\%$ information but the uncertainties in our
parameter estimates would be too large to have any meaningful
interpretation. The sum of all the eigenvalues $\lambda _i$
quantifies the total variance in the data and if we use the first
$M$ principal components then it encloses $r_M$ $\%$ of this
variance where
\begin{equation}
r_M=100 \frac{\sum_{i=1}^{M} \lambda _i}{\sum_{i=1}^{N} \lambda _i}.
\label{rm}
\end{equation}

\section{Results}

We use Markov chain Monte Carlo (MCMC) method (Metropolis-Hastings
MCMC chains) for estimating the parameters and their corresponding
errors. We show the best determined eigenmodes for the different
dataset combinations. In the plots the different parameter estimates
are found by allowing different number of dark energy evolution
parameters to vary. First we combine the supernovae measurements
with the CMB distance priors and resulting constraints are shown in
Figure \ref{sc1} and \ref{sc}. Then we add the BAO measurements to
this data set to see how they effect the parameter constraints.
Following \cite{jason} we employ the following trick when evaluating
the Fisher matix after the addition of BAO data: we keep $H(z)$ in
$d_z$ as fixed. If we don't do this we get spike like features in
our eigenfunctions. The results after addition of BAO data are shown
in figure \ref{scb1} and \ref{scb2}. For plotting convenience,
instead of plotting $\alpha_0$ we show the difference 0.734 -
$\alpha_0$ (the WMAP7 constraint on $\Omega_\Lambda$ in the
$\Lambda$CDM model is $\Omega_\Lambda \sim 0.734$). We summarise our
main results below:
\begin{itemize}
\item In the plots above we present the results obtained by
varying different number of principal components for each data
set. We observe that the constraint on the curvature parameter
$\Omega_k$ is robust against variation in the number of principal
component chosen for reconstruction. This indicates that the
curvature and dark energy evolution parameters have been
decorrelated. The results are consistent with a flat
Universe.

\item The addition of CMB priors is crucial. The analysis
using only the distance measurements resulted in a very wide allowed
range for $\Omega_k$ and the dark energy parameters.

\begin{figure}
\centering  
{\includegraphics[width=4in]{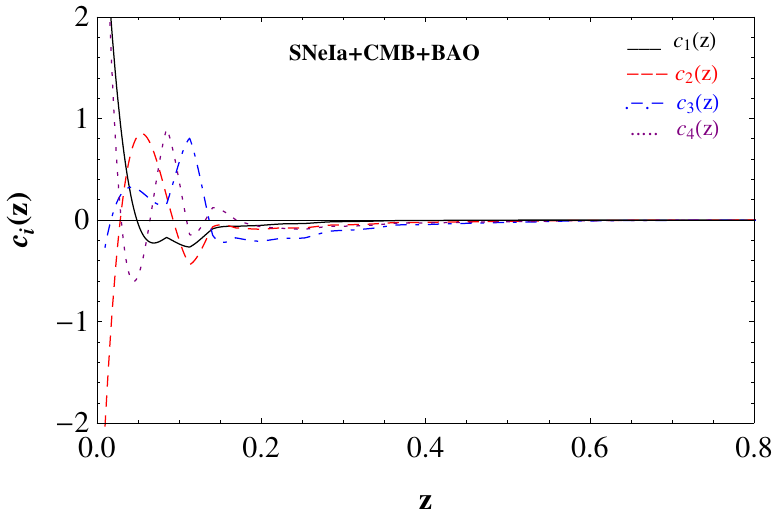}}
\caption{The panel shows the eigenmodes obtained
for the Supernovae+CMB+BAO data set. Solid (black), dashed (red),
dot-dashed (blue) and dotted (purple) curves correspond to the
first, second, third and fourth eigenmode respectively.}
\label{scb1}
\end{figure}

\begin{figure}
\centering  
\subfloat[Part
1][Dark energy parameters]{\includegraphics[width=3.0in]{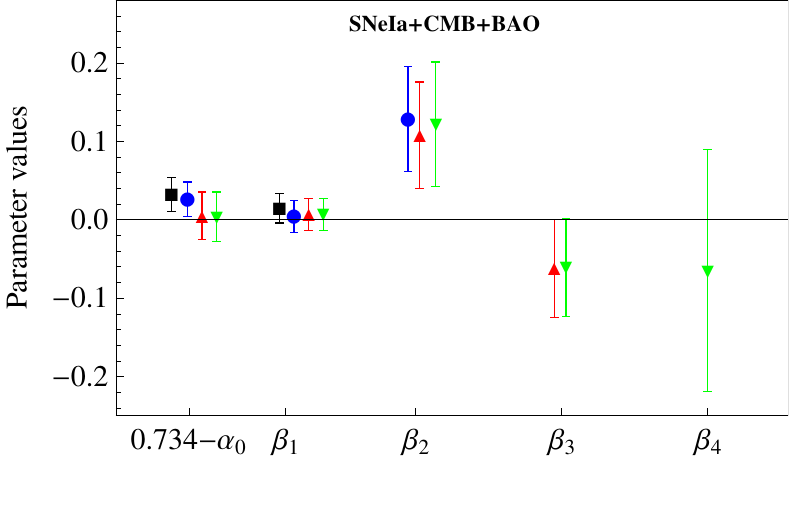}\label{scb2}}
~~~~~\subfloat[Part 1][Curvature parameter]{\includegraphics[width=3.2in]{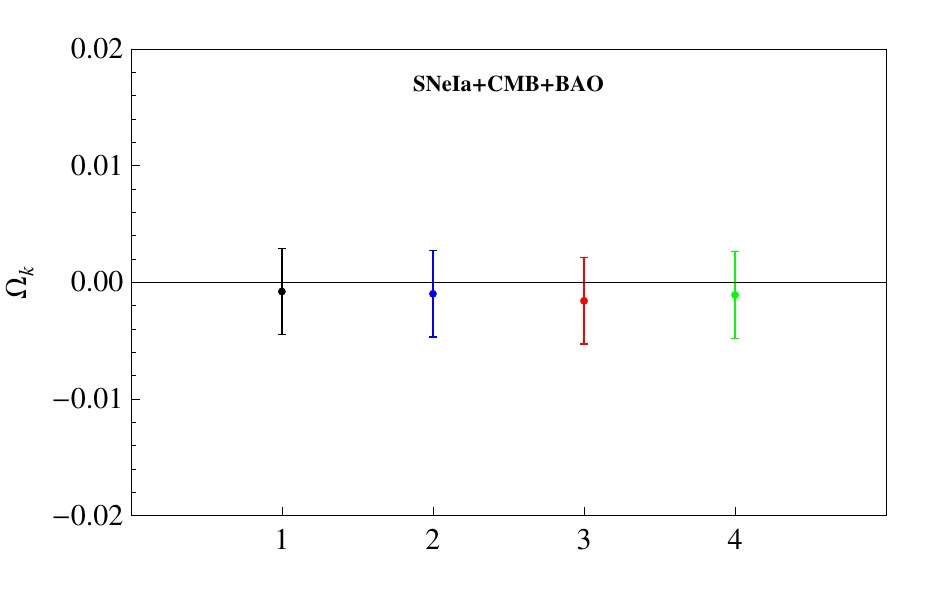} \label{kcb2}}
\caption{The left panel shows the estimates for the dark energy
parameters. Color scheme (from left to right):
Black/square (Case 1): $\alpha_0$ $\&$ $\beta_1$ allowed to vary,
Blue/circle (Case 2): $\alpha_0$, $\beta_1$ $\&$ $\beta_2$ allowed to
vary, Red/up arrow (Case 3): $\alpha_0$, $\beta_1$, $\beta_2$ $\&$
$\beta_3$ allowed to vary, Green/down arrow (Case 4): $\alpha_0$,
$\beta_1$, $\beta_2$, $\beta_3$ $\&$ $\beta_4$ allowed to vary. The right panel shows the curvature 
parameter estimates.}
\end{figure}

\item The constraints obtained from the supernovae-CMB combined measurements are consistent with a flat $\Lambda$CDM Universe. The first few principal components are consistent with zero. The reconstruction is done using upto 4 principal components. This amounts to incorporating almost 97$\%$ of the information (see \ref{rm}).

\item The addition of the BAO data to the supernovae-CMB data set improves the parameter estimates but also changes their value. The reconstruction is done using upto 4 principal components. This amounts to incorporating almost 94$\%$ of the information (see \ref{rm}). The coefficients of the first two eigenmodes are now slightly shifted from zero.
\end{itemize}

\section{Discussion}
Finding constraints on both the flatness of the Universe and the dark energy parameters, in models that allow dark energy evolution is difficult because of the geometric degeneracy. Weak assumptions on the possible evolution of dark energy can limit the allowed values of $\Omega_k$ and some simple parametrization of the equation of state is often assumed to put simultaneous constraints. But such assumptions/parametrization can introduce significant bias in the analysis and the results obtained. For example Clarkson et al., showed that the assumption of a flat universe leads to large errors in reconstructing the dark energy equation of state even if the true cosmic curvature is very small \cite{clark}.

In this work we have used a non-parametric method: Principal Component Analysis, to look for evidence for evolution in the dark energy density. The dark energy density is expressed as a sum of two terms: a constant term that accounts for the contribution that is redshift independent and an additional term constructed from the non constant density contribution. This later term is formulated using PCA so that all the parameters obtained have uncorrelated errors and a non constant amplitude of these modes would indicate dark energy evolution. The distance-redshift data alone cannot break the degeneracy between curvature and dark energy parameters. One can use growth data to remove this degeneracy \cite {Mort}. Also since high redshift distances, for example the distance to the last scattering surface, is sensitive to the curvature, one can use this measurement to find simultaneous constraints on $\Omega_k$ and dark energy parameters. In this work we have used the WMAP7 distance priors ($l_A$, $R$ and $z_*$). We used the latest supernovae data along with the CMB distance priors and found that it is consistent with a flat $\Lambda$CDM Universe. Later we incorporated the recent BAO data to see its effect on the parameter estimates. The constraints obtained on the non-constant modes from the addition of BAO data are slightly shifted from zero. The second principal component obtained in this case is not consistent with zero at 1$\sigma$. A possible deviation in dark energy equation of state from -1 was recently shown by the $WMAP$ team (see Fig. 10 \cite{wmap9}). Also, Gong-Bo Zhao et al., used a new non-parametric Bayesian method for reconstructing the evolution history of the equation-of-state of dark energy and found that the cosmological constant appears consistent with current data, but that a dynamical dark energy model which evolves from $w<-1$ at $z\sim0.25$ to $w > -1$ at higher redshift is mildly favored \cite{gong}. It is important to note that the result we obtained could be due to some unknown systematic effect and future BAO measurements would play key role in understanding the dark energy dynamics.

\section*{Acknowledgement}

Authors thank Jason Dick and Deepak Jain for discussions.
R.N. acknowledges support under CSIR - SRF scheme (Govt. of India). SJ acknowledges financial support provided by
Department of Science and Technology, India under project No. SR/S2/HEP-002/2008.

\end{document}